\documentclass[fleqn,usenatbib]{mnras}

\usepackage{newtxtext,newtxmath}

\usepackage[T1]{fontenc}

\DeclareRobustCommand{\VAN}[3]{#2}
\let\VANthebibliography\thebibliography
\def\thebibliography{\DeclareRobustCommand{\VAN}[3]{##3}\VANthebibliography}

\usepackage{graphicx}	
\usepackage{amsmath}	
\usepackage{overpic}	
\usepackage[table]{xcolor} 
\usepackage{multirow} 
\usepackage{siunitx} 

\definecolor{lightgray}{HTML}{E0E0E0}

\newcommand{\hypersec}[1]{Section~\ref{#1}} 
\newcommand{\hyperapp}[1]{Appendix~\ref{#1}} 
\newcommand{\hyperfig}[1]{Fig.~\ref{#1}} 
\newcommand{\hypertab}[1]{Table~\ref{#1}} 
\newcommand{\hypersubfig}[2]{Fig.~\hyperref[#1]{\ref*{#1}#2}} 
\newcommand{\hypereq}[1]{equation~(\ref{#1})} 
\newcommand{\hypereqalt}[1]{equation~\ref{#1}} 
\newcommand\edtai[1]{#1}

\newcommand\editj[1]{#1}
\newcommand\edita[1]{#1}

\newcommand\rev[1]{#1} 

\title[LISA Binaries]{Population synthesis predictions of the Galactic compact binary gravitational wave foreground detectable by LISA}

\author[J. McMillan et al.]{Jake McMillan,$^{1,2}$\thanks{E-mail: \href{mailto:jake.mcmillan@durham.ac.uk}{jake.mcmillan@durham.ac.uk}}
Adam Ingram,$^{1}$\thanks{E-mail: \href{mailto:adam.ingram@newcastle.ac.uk}{adam.ingram@newcastle.ac.uk}}
Cordelia Dashwood Brown,$^{3}$
Andrei Igoshev,$^1$
Matthew Middleton,$^{3}$\newauthor
Grzegorz Wiktorowicz,$^{4}$
and Simone Scaringi$^{5,6}$
\\
$^{1}$School of Mathematics, Statistics, and Physics, Newcastle University, Newcastle upon Tyne NE1 7RU, UK\\
$^{2}$Centre for Advanced Instrumentation, Department of Physics, Durham University, South Road, Durham, DH1 3LE, UK\\
$^{3}$School of Physics and Astronomy, University of Southampton, Southampton SO17 1BJ, UK\\
$^{4}$Nicolaus Copernicus Astronomical Center, Polish Academy of Sciences, Bartycka 18, 00-716 Warsaw, Poland\\
$^{5}$Centre for Extragalactic Astronomy, Department of Physics, Durham University, South Road, Durham, DH1 3LE , UK\\
$^{6}$INAF -- Osservatorio Astronomico di Capodimonte, Salita Moiariello 16, I-80131 Naples, Italy
}

\date{Accepted XXX. Received YYY; in original form ZZZ}

\pubyear{\the\year{}}

\begin{document}
\label{firstpage}
\pagerange{\pageref{firstpage}--\pageref{lastpage}}
\maketitle

\begin{abstract}
We use population synthesis modelling to predict the gravitational wave (GW) signal that the Laser Interferometer Space Antenna (LISA) will  detect from the Galactic population of compact binary systems. We implement a realistic star formation history with time and position-dependent metallicity, and account for the effect of supernova kicks on present-day positions. We consider all binaries that have a white dwarf (WD), neutron star (NS), or black hole primary in the present-day. We predict that the summed GW signal from all Galactic binaries will already be detectable 3 months into the LISA mission, by measuring the power spectrum of the total GW strain. \edita{We provide a simple publicly available code to calculate such a power spectrum from a user-defined binary population.} In the full 4 year baseline mission lifetime, we conservatively predict that $>2000$ binaries could be individually detectable as GW sources. We vary the assumed common envelope (CE) efficiency $\alpha$, and find that it influences both the shape of the power spectrum and the relative number of \editj{detectable} systems with WD and NS progenitors. In particular, the ratio of \editj{individually detectable} binaries with chirp mass $\mathcal{M} < M_\odot$ to those with $\mathcal{M} \geqslant M_\odot$ increases with $\alpha$. We therefore conclude that LISA \rev{may} be able to diagnose the CE efficiency, which is currently poorly constrained.
\end{abstract}




\section{Introduction}

Gravitational waves (GWs) are ripples in spacetime that propagate at the speed of light, generated when masses accelerate \citep[e.g.][]{2023PhRvX..13d1039A}. A key class of GW sources are compact objects, which are dense stellar remnants such as white dwarfs (WDs), neutron stars (NSs), or black holes (BHs). The field of GW astronomy became a reality in 2015 with the direct detection of GWs from a binary BH merger by the Laser Interferometer Gravitational Wave Observatory (LIGO; \citealt{2016PhRvL.116f1102A}). In the decade since, the LIGO, Virgo, and Kamioka Gravitational Wave Detector (KAGRA) network has detected an extensive catalogue of compact object binary mergers, including BH-BH, NS-NS, and BH-NS events \citep{2023PhRvX..13d1039A}. In the 2030s, the European Space Agency is set to launch the Laser Interferometer Space Antenna (LISA), a space-based GW detector designed to explore low-frequency GWs in the $0.1~\mathrm{mHz}$ to $1~\mathrm{Hz}$ range \citep{2015CQGra..32a5014M,2017arXiv170200786A}. Unlike current ground-based detectors, which target high-frequency signals, LISA will primarily target merging massive BHs, extreme mass ratio in-spirals, and other potential sources of GWs such as cosmic strings \citep{2023LRR....26....2A}. Beyond its primary targets, LISA will also detect a continuous stochastic GW hum from a symphony of stable compact binary systems throughout our Galaxy, each emitting GWs mainly at twice their orbital frequency, $f=2f_\mathrm{orb}$ \citep[e.g.][]{ 2020PhRvD.101l3021L, 2020ApJ...898...71B, 2021PhRvD.104d3019K, 2023MNRAS.524.2836V, 2024MNRAS.530..844K}.

Given the expected abundance of compact binaries, their collective signal in LISA will form a Galactic foreground. This foreground is an important source of noise for the transient events that form the primary science targets \citep{2017JPhCS.840a2024C}, but it also provides rich insights into the binary population in our own Galaxy. Many studies have concentrated on WD-WD binaries, which have long been expected to dominate the foreground signal \citep{2004PhRvD..70l2002B,2006CQGra..23S.809S, 2010ApJ...717.1006R,2017MNRAS.470.1894K,2022PhRvL.128d1101S,2020ApJ...901....4B,2024ApJ...963..100K,2025ApJ...981...66D}. Recently, a growing number of studies have been considering other populations. For example, \cite{2023MNRAS.525L..50S} considered binary systems consisting of a WD and a main sequence (MS) star, and many studies have considered NS-NS binaries \citep{2015MNRAS.448.1078Y,2019MNRAS.483.2615K,2020ApJ...892L...9A,2020MNRAS.492.3061L,2021MNRAS.502.5576K,2023MNRAS.521.2368S}, with \rev{others} additionally considering BH-BH and BH-NS systems \citep{Wagg+2022,2001A&A...375..890N,2018MNRAS.480.2704L,2020MNRAS.494L..75S}. NS-WD binaries are now also thought to be significant contributors to the GW signal \citep{2018PhRvL.121m1105T,2019MNRAS.484..698R,2024MNRAS.529.1886H}, and recently even triple systems have been considered \citep{2025arXiv250813264X}. Given that LISA will detect the sum of all GW emission, it is an important step to model full Galactic populations rather than isolating particular object classes \citep[e.g.][]{2020ApJ...898...71B,2023MNRAS.524.2836V}.

The evolved population of compact objects emitting GWs in the LISA bandpass is highly sensitive to the details of stellar and binary evolution theory, meaning that measurements of the Galactic foreground will provide unique constraints on currently poorly understood physics. Among the processes studied are supernova kicks \citep[e.g.][]{2024MNRAS.530..844K} and common envelope (CE) evolution \citep[e.g.][]{2025ApJ...981...66D}. \editj{Indeed, a large fraction of evolved compact binaries in synthetic populations have typically been through a CE phase, during which the primary star's envelope expands towards the end of its life and engulfs its companion \citep{1976IAUS...73...75P}. GWs could therefore provide a new means of probing CE physics across a large population of binaries.} A key poorly constrained quantity is the CE efficiency, which is the efficiency with which the binary orbital energy can be converted to kinetic energy with which to eject the envelope from the system. 

Here, \editj{we evolve MS binaries using the population synthesis code \textsc{cosmic} \citep{2020ApJ...898...71B} to build realistic synthetic populations of compact binaries, including WDs, NSs, and BHs}, and predict the resulting LISA Galactic foreground signal. We account for realistic star formation history including cosmic evolution of spatial density and metallicity within the Galaxy, as well as the effect of supernova kicks on the present-day position of stellar remnants. We consider both the detection of individual sources, the detection of the overall population as broadband noise, and the ``unresolvable signal'' left over after the subtraction of the individually detected sources. We simulate several different populations, varying the CE efficiency to study the effect it has on the predicted signal measurable by LISA. The remainder of this paper is structured as follows. \hypersec{sec:pop_synth} details the generation of a Galactic compact binary population, \hypersec{sec:GW_modelling} describes our methods for computing the GW signal, and \hypersec{sec:analyse_signal} presents our results. We discuss our results in \hypersec{sec:discussion} and make concluding remarks in \hypersec{sec:conclusion}.

\begin{table*}\renewcommand{\arraystretch}{1.4}
    \centering
    \caption{Summary of the three-component Milky Way model from \citet{Wagg+2022}. The semi-empirical SFH of each component is normalised to reproduce the present-day stellar mass distribution \citep{2015ApJ...806...96L}, with an equal mass split between the discs following \citet{2014ApJ...781L..31S}. \rev{Star formation in each component is restricted to its respective star-forming time interval, $\tau_\star$.} The radial and vertical profiles follow exponential forms with varying scale lengths $R_d$ and $z_d$. For the thin disc, $R_d$ evolves with time as $R_\mathrm{\exp}(\tau)$ defined in \citet[equation 4]{Wagg+2022}. Metallicity depends on the age-position-metallicity relation $ \big[\mathrm{Fe/H}\big](R,\tau)$ described in \citet[equation 7]{2018ApJ...865...96F}.}\label{tab:MW}
    \begin{tabular}{lcccccccc}
        \hline
        Component & $\tau_\rev{\star}/\mathrm{Gyr}$ & SFH Form & $M/\mathrm{M_\odot} \times 10^{10}$ & Radial Form & $R_d/\mathrm{kpc}$ & Vertical Form & $z_d/\mathrm{kpc}$ & Metallicity$/\mathrm{Z_\odot}$\\
        \hline
        Bulge       & $6-12$  & $(\tau - 6~\mathrm{Gyr})(\rev{12~\mathrm{Gyr} - \tau})^2$\hspace{-1em} & $0.9$  & \multirow{3}{*}{$\displaystyle \frac{R}{R_d^2}\exp\left( - \frac{R}{R_d}\right)$} & $1.5$ & \multirow{3}{*}{$\displaystyle \frac{1}{z_d}\exp\left( - \frac{|z|}{z_d}\right)$} & $0.2\phantom{0}$ & \multirow{3}{*}{$\displaystyle 10^{\displaystyle 0.977\big[\mathrm{Fe/H}\big]}  $} \\
        
        Thick Disc  & $8-12$  & \multirow{2}{*}{$\displaystyle \exp\left[ - \frac{(\tau_\mathrm{MW}-\tau)}{\tau_\mathrm{SFR}} \right]$} & $2.6$  & & $2.3$ & & $0.95$ & \\
        
        Thin Disc   & $0-8\phantom{1}$ & & $2.6$  & &  $R_\mathrm{exp}$ & & $0.3\phantom{0}$ & \\
        
        Total & -- & -- & $6.1$  & -- & -- & -- & -- & --\\
        \hline
    \end{tabular}
\end{table*}

\section{Generating a Galactic compact binary population}\label{sec:pop_synth}

\edtai{Here we describe our methodology to simulate a synthetic present-day Galactic population of binaries.}

\subsection{Initial stellar population}
\label{sec:init_stellar_pop}

\edtai{We initialise a population of $10^{10}$ zero-age MS (ZAMS) binaries.} We randomly draw primary masses, $m_1$, from a \citet{2001MNRAS.322..231K} initial mass function (IMF), and we assign secondary masses, $m_2$, by drawing the mass ratio $q=m_2/m_1$ from a uniform distribution, with $q_\mathrm{min} < q < 1$. Here, $q_\mathrm{min}$ ensures the secondary's pre-MS lifetime does not exceed the primary's full lifetime if it were evolved in isolation, \editj{preventing scenarios where the primary star dies before the secondary even forms as a MS star.} \edtai{Following \citet{2012Sci...337..444S}, we sample orbital periods $P$ and eccentricities $e$ from power law distributions, with probability densities $\propto \log_{10}(P/\mathrm{day})^{-0.55}$ in the range $0.15 < \log_{10}(P/\mathrm{day})<5.5$, and $\propto e^{-0.45}$ in the range $(0,1]$.}

\edtai{\editj{By} assuming a binary fraction of $50\%$ \citep{2001MNRAS.322..231K, 2011MNRAS.417.1684M}, the $10^{10}$ binaries in our sample have a total stellar mass (single stars plus binaries) of $6.1 \times 10^9~M_\odot$; \editj{we note that more recent work (e.g. \citealt{2017ApJS..230...15M}) finds higher binary fractions for massive stars, but $50\%$ is reasonable for the lower mass stars dominating the GW foreground.} This \editj{mass} is an order of magnitude less than the total stellar mass in the Milky Way \citep{2015ApJ...806...96L}. We therefore scale all of our results by a factor of 10 to extrapolate our sample up to the true size of the Milky Way. We chose our sample size to be as large as possible within the bounds of computational feasibility in order to capture rare systems.}

\subsection{Milky Way model}
\label{sec:milky_way} 

\edtai{For each binary in our sample, we assign a} birth time, an initial Galactic position $(R,\theta, z)$, and a metallicity according to the three-component Milky Way model of \citet{Wagg+2022}, which is based upon e.g. \cite{2018ApJ...865...96F,2016ApJ...823...30B,2019MNRAS.490.4740B,2011MNRAS.414.2446M}. This model is summarised in \hypertab{tab:MW} (also see Fig.~1 of \citealt{Wagg+2022}). \edtai{We first assign a birth time for each binary by drawing from the total star formation history (SFH) shown in \hypersubfig{fig:Population_distributions}{c}. This is a sum of the bulge, thin disc, and thick disc components \rev{(\hypersubfig{fig:Population_distributions}{c}; dashed lines)}.} The bulge SFH is modelled with $\mathcal{B}(2,3)$ beta distribution, shifted to \rev{only} form stars between $12-6~\mathrm{Gyr}$ ago. Both discs follow an exponentially decaying SFH with a characteristic timescale $\tau_\mathrm{SFR}=6.8~\mathrm{Gyr}$ \edtai{and a total Milky Way age of $\tau_{\rm MW}=12$ Gyr}. We assume the thick disc formed between $12-8~\mathrm{Gyr}$ ago, after which star formation transitioned to the thin disc until the present-day.

\edtai{After selecting the birth time of a binary, we assign it to a Milky Way component (bulge, thick disc, or thin disc) based on the relative contribution of each component to the total SFH at the birth time. We then draw birth radial and vertical Galactic positions from exponential functions with parameters depending on the component (see \hypertab{tab:MW}). The radial scale length of the exponential, $R_d$, is constant for the bulge and the thick disc, but instead evolves with lookback time $\tau$ for the thin disc as $R_d = R_\mathrm{exp}(\tau)=4[1-0.3(\tau/8~\mathrm{Gyr})]~\mathrm{kpc}$, which accounts for its inside-out growth.} We sample the birth Galactic azimuthal angle uniformly across $[0,2\pi)$ radians, and assign metallicity using the age-position-metallicity relation of \citet[][equation 7]{2018ApJ...865...96F}.

\subsection{Population evolution}
\label{sec:pop_evolution}

\edtai{We evolve each binary from its selected birth time to present-day using the} \textsc{cosmic} population synthesis code \rev{(v3.6.1; \citealt{2025zndo..15164778C, 2020ApJ...898...71B})}. \textsc{cosmic} combines the Single Star Evolution model for single stars \citep{2000MNRAS.315..543H} with a modified version of the Binary Star Evolution code for binaries \citep{2002MNRAS.329..897H}. It implements an extensive set of population synthesis parameters, whose defaults for this work are summarised in \hyperapp{appendix:cosmic_parameters} and detailed in the \textsc{cosmic} documentation\footnote{\url{https://cosmic-popsynth.github.io/docs/}}. During evolution, \textsc{cosmic} accounts for supernova kicks, orbital shrinking due to GW emission, mass transfer, tides, and other relevant binary evolution processes according to standard prescriptions implemented in population synthesis (for details see \citealt{2020ApJ...898...71B}).

\begin{figure}
    \centering
    \includegraphics[width=\columnwidth]{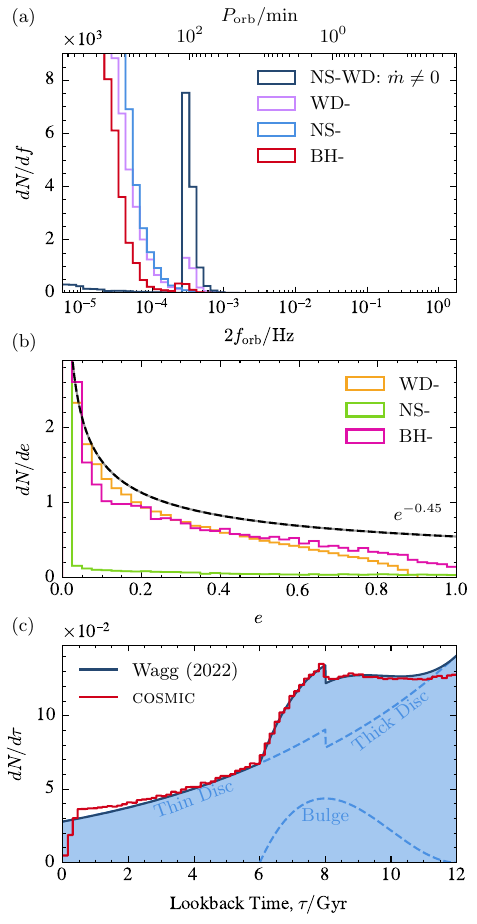}
    \vspace{-1.5\baselineskip}
    \caption{Properties of the generated compact binary population. (a) Orbital period distribution within LISA's band, split into primary type (i.e. WD, NS, or BH) and accreting ($\dot{m}\neq 0$) NS-WD systems. The majority of the population (not plotted) lies outside LISA's band towards $10^{-10}~\mathrm{Hz}$. (b) Eccentricity distribution, split by primary type\rev{, compared with the input model from \citet[dashed line]{2012Sci...337..444S}}. Given there are nearly $5\times 10^6$ circular binaries, their peak at $e=0$ has been truncated for visibility. (c) Star formation history of the full compact binary population compared with the theoretical SFH from \citet{Wagg+2022}, showing fewer compact binaries from very old and very recent times relative to the model.}
    \label{fig:Population_distributions} 
\end{figure}

We filter the evolved population to retain only binaries containing at least one WD, NS, or BH. For subsequent analysis, we redefine the primary: if only one star is compact, it is taken as the primary; if both are compact, the more massive object is chosen. The final population \edtai{evolved with the default parameters (see \hyperapp{appendix:cosmic_parameters})} contains approximately $1.5\times10^8$ WD-primary, $3.3\times10^6$ NS-primary, and $2.2\times10^6$ BH-primary binaries, giving a total of nearly $1.6\times 10^8$ compact object binaries with orbital periods ranging from $2$~mins to $20{,}000$~years. These numbers are scaled by a factor of 10 to be more comparable to the full Galactic compact binary population (see \hypersec{sec:init_stellar_pop}). The distribution of orbital periods for this population is shown in \hypersubfig{fig:Population_distributions}{a} across the LISA band, with the majority of binaries well within the LISA frequency band (i.e. where the sensitivity is highest, see \hypersec{sec:intrumental_noise}) being accreting NS-WD systems. The smaller WD- peak \edtai{at} $\sim 3\times 10^{-4}~\mathrm{Hz}$ mainly contains \edtai{WD-MS binaries, meaning that these systems correspond to} the cataclysmic variables studied by \citet{2023MNRAS.525L..50S}. 

\editj{Our present-day population can be further classified into circular ($e=0$) and eccentric ($e>0$) orbits}, numbering roughly $5\times 10^7$ and $11 \times 10^7$ systems, respectively. Here, the orbital eccentricity, $e$, ranges from $0$ (perfectly circular) to $1$ (parabolic). The distribution of eccentricities is shown in \hypersubfig{fig:Population_distributions}{b}. WD- and BH-primary binaries show an approximately \rev{exponential} decrease in number with increasing eccentricity (\rev{following the input model; black dashed line}), whereas NS-primary systems have a nearly uniform distribution of $e$ aside from a huge peak at $e=0$. \rev{We attribute the flat NS–primary distribution and sharp peak at $e=0$ to strong NS natal kicks (see \hypersec{sec:kicks}), which disrupt most NS-forming binaries. Systems which remain bound are those that received comparatively small kicks and were already circularised by prior mass transfer. WD- and BH-primary systems show weaker deviations from the input distribution because any natal kick is imparted by the companion NS rather than the primary.} \edtai{We plot the} reconstructed SFH \edtai{of our filtered binary population in \hypersubfig{fig:Population_distributions}{c} (red histogram). We see that, compared with the input model, there are} fewer surviving compact binaries from very early and very late times, consistent with many older systems having merged and many younger systems not yet having evolved into compact objects.

\subsection{Kicks and spatial distribution}
\label{sec:kicks}


During evolution, \textsc{cosmic} tracks velocities imparted to binaries by both Blaauw \citep{1961BAN....15..265B} and natal (e.g. see \citealt{2005MNRAS.360..974H}) supernova kicks \rev{through a combined centre-of-mass kick}. Blaauw kicks arise from mass loss during the supernova explosion \edita{(}or direct collapse\edita{) shifting the binary centre of mass}, while natal kicks are \edita{caused by asymmetries in the supernova explosion.} For core-collapse supernovae, \edita{we set \textsc{cosmic} to draw} kick magnitudes from a Maxwellian distribution with dispersion $\sigma = 265~\mathrm{km\,s^{-1}}$, following \citet{2005MNRAS.360..974H}. Electron-capture and ultra-stripped supernovae receive a lower dispersion of $20~\mathrm{km\,s^{-1}}$ \citep{2020ApJ...898...71B}, while BH kicks are modulated by fallback following \citet{2012ApJ...749...91F}.
We account for the effect of these supernova kicks on the position of \edtai{each binary by first selecting an isotropically assigned direction for the kick (the magnitude is taken from the \textsc{cosmic} run), then} evolving the binary in the Milky Way potential \edtai{from the birth position to the time of the supernova to a new position in the present-day. For binaries that experience two supernovae, we perform this calculation twice, evolving the binary position from the time of the first supernova to the time of the second supernova, and finally to present-day.} To do this, we use \textsc{galpy} \citep{2015ApJS..216...29B} assuming a composite Milky Way potential (specifically an implementation of the ``MilkyWayPotential2022'' from \citealt{gala}). Our procedure follows \cite{2024MNRAS.530..844K}, where a detailed description of the assumptions can be found. For computational efficiency, we only integrate binaries receiving a kick with orbital frequencies above $20~\mathrm{\mu Hz}$ (corresponding to GW frequencies $>40~\mathrm{\mu Hz}$ and, importantly, within LISA's band). \rev{We note that the recent \textsc{cogsworth} code \citep{2025ApJS..276...16W} extends \textsc{cosmic} by incorporating full SFHs and performing an orbital-integration procedure similar to our approach.}

\edtai{For each Galactic population we evolve with \textsc{cosmic}, we select multiple different realisations of the present-day binary positions, so that we can later average our results to smooth out Monte Carlo noise.} For non-kicked binaries, we reselect azimuthal angle and height for each realisation, whereas each binary keeps the same radial position for every realisation because this is tied to metallicity. For each kicked binary, we retain the same $R$ and $|z|$ because these are set by the orbital integration \edtai{(i.e. to randomise $z$, we would need to run \textsc{galpy} for each new realisation, which would be prohibitively computationally expensive)}, but we redraw the azimuthal angle and randomly flip the sign of $z$ to place the kicked binary above or below the Galactic plane.

\begin{figure}
    \centering
    \includegraphics[width=\columnwidth]{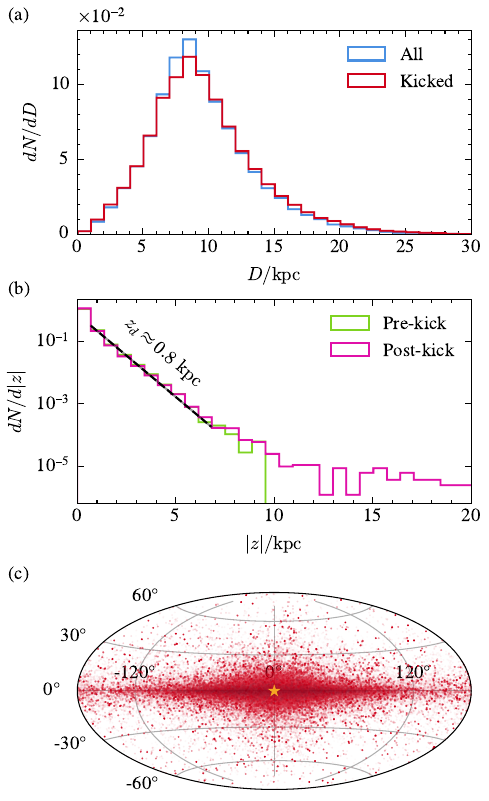}
    \vspace{-1\baselineskip}
    \caption{(a) Distance distribution (from Earth) of the generated compact binary population, showing a peak near $8~\mathrm{kpc}$ at the Galactic bulge. Kicked binaries show a deficit near the Galactic centre, as some systems are ejected to larger distances. \rev{(b) Height above the Galactic plane, $|z|$, for the population pre-kick and post-kick. Both follow an effective scale height of $z_d \approx 0.8~\mathrm{kpc}$ at low $|z|$, but post-kick systems develop a high-$|z|$ tail extending beyond $10~\mathrm{kpc}$.} (c) Sky map of the compact binaries in Galactic coordinates, showing the concentration of sources in the bulge near the Galactic centre (star) and a disc-like structure.}
    \label{fig:Population_Distance}
\end{figure}

From the resulting present-day 3D binary positions, we calculate the distance to Earth by placing Earth at $(8~\mathrm{kpc}, 0, 0)$ in the Galaxy. The resulting distance distribution for one realisation is shown in \hypersubfig{fig:Population_Distance}{a}, peaking near $8~\mathrm{kpc}$ due to the bulge, with extended tails from the discs. Focusing on kicked binaries alone reveals a relative decrease near the Galactic centre, likely because some systems born there were ejected out of the Milky Way. \rev{\hypersubfig{fig:Population_Distance}{b} shows the height-above-plane distribution for this realisation before (pre-kick) and after (post-kick) orbital integration. At low $|z|$, both follow the vertical exponential profile in \hypertab{tab:MW} with an effective scale height of $z_d \approx 0.8~\mathrm{kpc}$ (black dashed line), consistent with a weighted contribution of the three components' individual profiles. At larger heights, the effect of kicks becomes apparent: pre-kick binaries remain confined within $|z|\lesssim10~\mathrm{kpc}$, whereas post-kick systems develop a high-$|z|$ tail that reaches well past $10~\mathrm{kpc}$ for the most strongly kicked or centrally born systems.} A sky map \edtai{of the same realisation} in Galactic coordinates is shown in \hypersubfig{fig:Population_Distance}{c}.



\subsection{Common envelope}

A CE phase occurs when one star's envelope expands towards the end of its life to engulf its companion \citep{1976IAUS...73...75P}. As the companion spirals inward due to friction, orbital energy is transferred to the envelope. If sufficient energy is deposited, the envelope will be ejected, leaving behind a tighter post-CE phase binary in which mass transfer may subsequently occur. If not enough energy is deposited, the envelope remains and the two stars ultimately merge.
There are two basic approaches to model the CE stage: (1) energy based and (2) angular momentum based. In the first approach it is assumed that the orbit shrinks and a fraction of released gravitational energy is used to unbind the stellar envelope. Similarly in the second approach, a fraction of orbital angular momentum is used to unbind the stellar envelope. 

The energy-based prescription is often called the $\alpha$ formalism for CE. Its key ingredient is CE efficiency defined as \citep{1984ApJ...277..355W,2014MNRAS.442.1980Z} 
\begin{equation}
    \alpha = \frac{E_\mathrm{bind}}{\Delta E_\mathrm{orb}},
\end{equation}
where $E_\mathrm{bind}$ is the binding energy of the envelope and $\Delta E_\mathrm{orb}$ is the change in orbital energy from the reduction of the orbital separation during the CE phase. The envelope's binding energy $E_\mathrm{bind}$ depends on its mass, \edtai{the Roche Lobe radius of the evolved star}, and internal structure through a parameter $\lambda$. \edtai{Our \textsc{cosmic} runs employ a value of $\lambda$ that depends on the evolutionary state of the star (as described in Appendix A of \citealt{2014A&A...563A..83C})}. The orbital energy change $\Delta E_\mathrm{orb}$ between the initial and final binary separations \edtai{is calculated} using the virial theorem.

The angular momentum based prescription is usually called the $\gamma$ formalism for CE \citep{2000A&A...360.1011N}. This formalism suggests that the first CE leading to double WD formation proceeds differently in comparison to the second CE. Recently, \citet{Nelemans2025AA} showed that the first mass transfer indeed proceeds differently from the second one, but $\gamma$-CE does not describe it precisely. \edita{Here, we only employ the $\alpha$ formalism because the $\gamma$ formalism is not implemented in \textsc{cosmic}.}

\subsection{Population variation}\label{sec:pop_var}

To explore the impact of varying population synthesis parameters, we repeat the \textsc{cosmic} method using different values of the CE efficiency, $\alpha$, with all other parameters held constant. \editj{Estimates of $\alpha$ vary (indeed, it is likely that $\alpha$ is not constant; \citealt{1995MNRAS.273..146R}), with $\alpha = 1$ fairly typical in literature (e.g. \citealt{2001A&A...368..939N}), although may be as low as $\alpha \sim 0.2$ \citep{2010A&A...520A..86Z, 2022MNRAS.513.3587Z} to as high as $\alpha \sim 5$ \citep{2019ApJ...883L..45F}.} \edtai{Thus, our default runs employ $\alpha=1$, and we explore a range of values from $\alpha=0.2$ to $\alpha=5$.}

\begin{figure*}
    \centering
    \includegraphics[width=\textwidth]{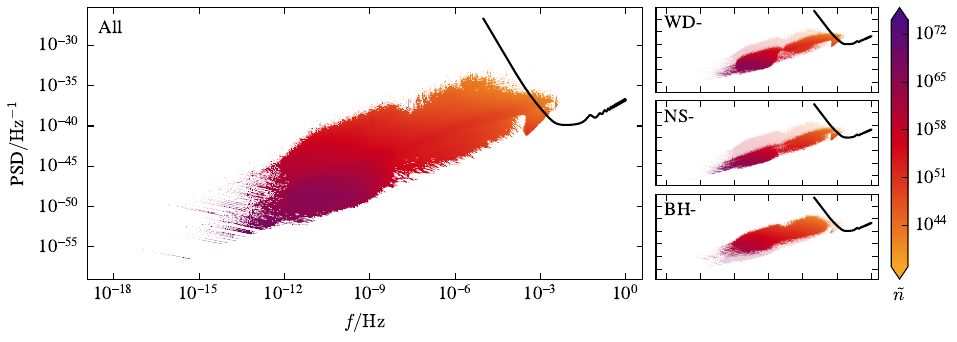}
    \vspace{-1.5\baselineskip}
    \caption{Number density of \rev{GW signals from} binaries per \editj{unit log-frequency-log-power area}, projected against LISA's sensitivity curve. The right panels split the population by primary type (i.e. WD, NS, or BH), showing most BH-primary systems lie well below the curve. WD-primary binaries roughy divide into two groups: lower-frequency eccentric systems and higher-frequency circular systems. A detectable subset of NS-primary binaries clusters near $10^{-3}~\mathrm{Hz}$.}
    \label{fig:Eccentric_Number_Density} 
\end{figure*}

\section{Gravitational wave modelling}\label{sec:GW_modelling}

LISA will consist of three satellites arranged in an equilateral triangle with a mean separation of $2.5$ million km, following a heliocentric orbit trailing behind Earth \citep{2017arXiv170200786A}. It will detect GWs by precisely measuring changes in the distances between its satellites using laser interferometry. The measured strain as a function of time is $x(t) = n(t) + h(t)$, where $h(t)$ is the signal (i.e. movement of the arms due to GWs passing by) and $n(t)$ is instrumental noise (i.e. movement of the arms for any other reason). The power spectral density (PSD) of the strain is $S(f)=|X(f)|^2$, where $X(f)$ is the Fourier transform of $x(t)$. In this section, we detail how we calculate the PSD of both the instrumental noise and of the signal.

\subsection{Instrumental noise}\label{sec:intrumental_noise}

\edtai{The PSD of LISA's instrumental noise (the LISA sensitivity curve) is shown in \hyperfig{fig:Eccentric_Number_Density} (black line), calculated from \citet[][equations 1, 10, and 11]{Robson_2019}. Note that this is purely instrumental noise from non-astrophysical effects, and thus does not include confusion noise from unresolved astrophysical sources (which we will discuss in Section \ref{sec:analyse_signal}). We see that the noise is lowest in the approximate range $1~\mathrm{mHz}$ to $1~\mathrm{Hz}$.}

LISA's noise curve is shaped by different contributions across the frequency spectrum \citep{2003PhRvD..68j2002S}. At low frequencies, acceleration noise from forces on the test masses and radiation pressure from the laser limit sensitivity. At high frequencies, optical metrology noise dominates due to quantum fluctuations in the laser and Poisson noise from limited photon counts. The oscillations at high frequencies occur due to LISA's arm length affecting the response to signals with shorter wavelengths \citep{2000PhRvD..62f2001L}. There is a trade-off in beam power: increasing it reduces high-frequency Poisson noise but increases low-frequency acceleration noise due to increased photon pressure. To achieve optimal sensitivity for its targets, LISA's laser power, test mass shielding, and spacecraft configuration are carefully tuned.

\subsection{Circular binary emissions}\label{sec:circ_bin_emission}

For the Galactic binaries detectable by LISA, the orbital period is typically stable on timescales of years, meaning that their GW waveforms are much easier to calculate than those of the mergers that ground based detectors are sensitive to \citep[e.g.][]{2006PhRvL..96k1101C}. For stable circular binary orbits, the strain is simply a sine wave function of time with frequency $f=2/P_\mathrm{orb}$, where $P_{\rm orb}$ is the orbital period. The amplitude, averaged over LISA's sky position and binary inclination, is well approximated by \citep{2004PhRvD..70b2003K,2016gwdw.book....9W}
\begin{equation}
    \mathcal{A} = \frac{4(G\mathcal{M})^{5/3}}{c^4 D} \sqrt{\frac{4}{5}} \left(\frac{2\pi}{P_\mathrm{orb}}\right)^{2/3}, \label{eq:amplitude_circular}
\end{equation}
where $G$ is the gravitational constant, $c$ is the speed of light, $D$ is the distance to the binary, and $\mathcal{M}$ is the chirp mass
\begin{equation}
    \mathcal{M} = \left( \frac{m_1^3 m_2^3}{m_1+m_2}\right)^{1/5}. \label{eq:chirp_mass}
\end{equation}

As a GW from a single binary crosses the detector, its power at frequency $f=2/P_{\rm orb}$ is
\begin{equation}
    P = \mathcal{A}^2 T_\mathrm{obs}\label{eq:perfect_power},
\end{equation}
where $T_\mathrm{obs}$ is the observing time, which we take by default to be the nominal LISA mission duration of $4~\mathrm{yr}$ \citep{2017arXiv170200786A}. The corresponding signal-to-noise ratio (SNR) of a binary is
\begin{equation}
    \rho = \mathcal{A}\sqrt{\frac{T_\mathrm{obs}}{S_\mathrm{noise}(f)}},\label{eq:SNR}
\end{equation}
where $S_\mathrm{noise}(f)$ is the LISA noise PSD evaluated at the binary's GW frequency $f=2/P_\mathrm{orb}$. Sources are typically considered to be individually detectable if they have $\rho > 7$ (as in e.g. \citealt{2023PhRvD.107f4021L}). \edtai{However, the above equation only represents the SNR above the \textit{instrumental} noise. In reality, each source also needs to be distinguished from the summed signal from all other sources in the sky (i.e. confusion noise), which we consider in Section \ref{sec:resolvable_binaries}.}


\subsection{Eccentric binary emissions}\label{sec:eccentric_binaries}

As discussed in \hypersec{sec:pop_evolution}, around $70\%$ of the population is eccentric. \rev{While the circular orbit approximation provides a useful foundation, we must account for eccentricity to capture the full GW signal.} Eccentric binaries emit GWs across a broad spectrum of harmonic frequencies, $f_\eta = \eta/P_\mathrm{orb}$, where $\eta=1,2,\dots$ is the harmonic number. In contrast to circular systems, which emit only at the second harmonic, eccentric binaries distribute their GW power across multiple harmonics, producing this broader spectrum. For a stable eccentric binary, the instantaneous GW amplitude at the $\eta^\text{th}$ harmonic, averaged over LISA's sky position and binary inclination, is 
\begin{equation}
    \mathcal{A}_\eta = \frac{4(G\mathcal{M})^{5/3}}{c^4 D} \sqrt{\frac{4}{5}} \left(\frac{2\pi}{P_\mathrm{orb}}\right)^{2/3} \frac{\sqrt{g(\eta,e)}}{(\eta/2)}, \label{eq:amplitude_eccentric}
\end{equation}
where $e$ is the orbital eccentricity and the factor $\sqrt{g(\eta,e)}(\eta/2)^{-1}$ describes the relative amplitude of each harmonic compared to the circular amplitude. The function $g(\eta,e)$ is defined in \citet{1963PhRv..131..435P} as
\begin{equation}
\begin{split}
    g(\eta, e) = & ~ \frac{\eta^{4}}{32} \bigg\{ \left(1-e^{2}\right)\Big[J_{\eta-2}(\eta e)-2 J_{\eta}(\eta e)+J_{\eta+2}(\eta e)\Big]^{2}\\
    & + \Big[ J_{\eta-2}(\eta e) -2 e J_{\eta-1}(\eta e) + \frac{2}{\eta} J_{\eta}(\eta e) \\ & + 2 e J_{\eta+1}(\eta e)-J_{\eta+2}(\eta e)\Big]^{2} + \frac{4}{3 \eta^{2}}\Big[J_{\eta}(\eta e)\Big]^{2} \bigg\} ,
\end{split}
\end{equation}
where $J_n(x)$ is the Bessel function of the first kind of order $n$ with argument $x$. In the $e=0$ limit, every harmonic vanishes except for the second, with $\sqrt{g(2,0)}(2/2)^{-1}=1$, returning to the circular result as in \hypereq{eq:amplitude_circular} at frequency $f_2=2/P_\mathrm{orb}$. The corresponding SNR of an eccentric binary is
\begin{equation}
    \rho = \sqrt{\sum_{\eta = 1}^\infty \mathcal{A}_\eta^2 ~ \frac{T_\mathrm{obs}}{S_\mathrm{noise}(f_\eta)}}, \label{eq:SNR_eccentric}
\end{equation}
where $S_\mathrm{noise}(f_\eta)$ is the LISA noise PSD evaluated at the binary's $\eta^\text{th}$ harmonic frequency. 

In theory, each eccentric binary emits GWs at an infinite series of harmonics. However, for computation, we truncate the sum at an empirical harmonic cut-off, $\eta_\mathrm{c}$, chosen such that harmonics with relative amplitudes $<0.1$ (or $<1\%$ in power) are excluded
\begin{equation}
    \eta_\mathrm{c}(e) = \min\left\{ \left\lceil 3+ \left( 1 - e + \frac{e^2}{5} - \frac{e^{10}}{10}\right)^{-3} \right\rceil,~ 331\right\}.\label{eq:harmonic_cutoff}
\end{equation}



\rev{\hyperfig{fig:Eccentric_Number_Density} summarises the GW emission from our synthetic Galactic compact binary population, generated from the default parameters. Assuming all binaries to have a stable orbital period, we calculate the power in each harmonic of each source and the corresponding harmonic frequency.} We plot the number density of \rev{harmonics} per unit log-frequency-log-power area, projected on the LISA sensitivity curve, highlighting where \rev{signals} cluster across frequency-power space. \rev{We note that, whereas circular sources only emit GW power at the second harmonic ($\eta=2$), eccentric sources emit over several harmonics (with the total number of harmonics we consider defined by \hypereqalt{eq:harmonic_cutoff}). The diagonal features at lower frequencies reflect individual eccentric binaries whose GW power is distributed across multiple harmonics.}

At low frequencies, all \rev{signals} lie well below the noise curve and are undetectable. This region is dominated by BH-primary and eccentric WD-primary binaries, which tend to have longer orbital periods and weak GW amplitudes. As frequency increases, the number density decreases, but GW amplitude grows. A subset of the NS-primary and circular WD-primary binaries with sufficient amplitudes to be detected are clustered near $10^{-3}~\mathrm{Hz}$. Despite the larger number of WD-primary systems, they do not totally dominate.

\begin{figure*}
    \centering
    \includegraphics[width=\textwidth]{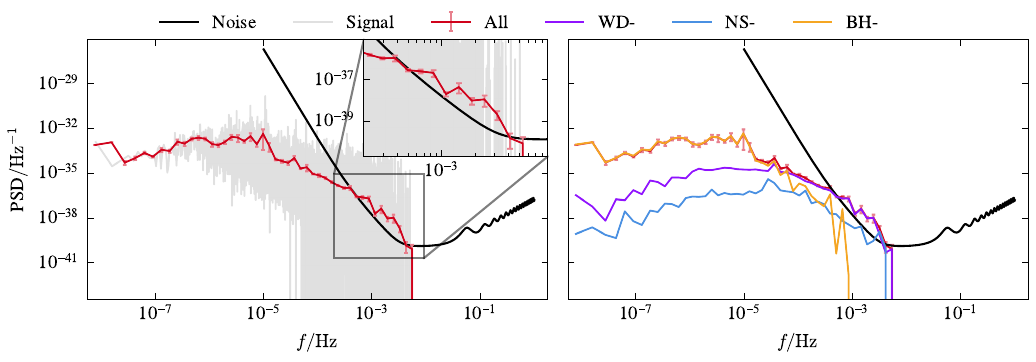}
    \vspace{-1.5\baselineskip}
    \caption{(Left) PSD as a function of GW frequency for compact object binaries, \rev{with the full contribution from eccentric harmonics included}. The raw, noisy signal is smoothed through geometric re-binning ($c_0=1.3$), shown with $1\sigma$ error bars. A detectable signal is clearly above the LISA noise curve. (Right) Comparison of GW PSDs split by primary type (WD, NS, or BH). BH systems dominate at low frequencies, whereas WD and NS systems govern the detectable higher frequencies within LISA's band.}
    \label{fig:Circular_DFT_Wide}
\end{figure*}

\subsection{Calculating the total signal}\label{sec:DFT_analysis_sub}



\rev{To compute the combined GW PSD from an ensemble of $\mathcal{N}_\mathrm{bin}$ binaries, we sum the contributions from all harmonics emitted by all systems. The total number of terms in this harmonic sum, $\mathcal{N}_\mathrm{har}$, would simply be equal to $\mathcal{N}_\mathrm{bin}$ if all the binaries were circular; but it is much larger for our population because the eccentric sources emit GWs at many harmonics. The strain of the $n^{\rm th}$ harmonic is}
\begin{equation}
    h_n(t_k) = \mathcal{A}_n \cos(2\pi f_n t_k - \phi_n),\label{eq:nth_strain}
\end{equation}
\rev{where $\mathcal{A}_n$ is the harmonic amplitude (\hypereqalt{eq:amplitude_eccentric}), $f_n$ is the harmonic frequency, and $\phi_n$ is the orbital phase. All harmonics of a given binary share the same phase, which is drawn uniformly random across $[0,2\pi)$.} This \rev{signal} is sampled at discrete times $t_k = k\,dt$, where $k=1,2,\dots,N$ and $T_\mathrm{obs}=N\,dt$ is the observing time. We use $dt=0.5~\mathrm{s}$ for our calculations, which sets a maximum (Nyquist) frequency of $1~\mathrm{Hz}$. The discrete Fourier transform of equation (\ref{eq:nth_strain}), normalised such that the integral of the PSD over all positive frequencies is equal to $\mathcal{A}_n^2$, is
\begin{equation}
    H_n(f_j) = \frac{1}{\sqrt{df}} \Theta_{jn} \mathcal{A}_n ~ \mathrm{e}^{i\phi_n},
\end{equation}
where $f_j = j\,df$ is the $j^\text{th}$ Fourier frequency, $df=1/T_\mathrm{obs}$ is the frequency resolution, and $\Theta_{jn}=1$ for \rev{harmonics} with $-df/2 \leq f_n - f_j < df/2$, and zero otherwise (see Appendix \ref{appendix:DFT} for a derivation). The power from a single \rev{harmonic}, $P_n(f_j) = |H_n(f_j)|^2$ is therefore a top hat function with width $df$ and height $\mathcal{A}_n^2/T_{\rm obs}$ (and thus the integral over all positive frequencies is $\mathcal{A}_n^2$, as specified). We therefore see that, the longer LISA observes for, a given GW \rev{harmonic} will be detected over an increasingly narrow frequency range with an increasing peak power.

The total GW strain is the sum of all \rev{harmonics across all binaries}, and thus we calculate the total signal PSD of our population as
\begin{equation}
    P(f_j) = |H(f_j)|^2 = \frac{1}{df} \Bigg| \sum_{n=1}^{\rev{\mathcal{N}_\mathrm{har}}} \Theta_{jn} \mathcal{A}_n ~ \mathrm{e}^{i\phi_n} \Bigg|^2.
\end{equation}
For very large $T_\mathrm{obs}$, each \rev{harmonic} falls into its own separate frequency bin, reducing the above to a quadrature sum, as is assumed by \citet{2023MNRAS.524.2836V}. \edtai{We then scale the PSD by the factor of 10 discussed in Section \ref{sec:init_stellar_pop} so that it represents} the expected total Galactic signal. Because the raw PSD is extremely noisy, we apply geometric re-binning, grouping Fourier frequencies into bins of geometrically increasing size (see e.g. \citealt{2012PhDT........51I}). The resulting raw and smoothed PSDs in \hyperfig{fig:Circular_DFT_Wide} (left) show a clearly detectable signal above LISA's noise curve, with the geometric re-binning parameter set to $c_0=1.3$. \hyperfig{fig:Circular_DFT_Wide} (right) shows the same summed PSD, broken down into populations of binaries with WD, NS, and BH primaries. BH systems dominate the low-frequency regime, whereas WD and NS systems dominate the high frequencies\edtai{, where they contribute a detectable signal}. \rev{BH systems dominate low frequencies because GW-driven inspiral scales as $df/dt \propto \mathcal{M}^{5/3} f^{11/3}$ \citep{1964PhRv..136.1224P}: heavier binaries evolve fastest, and the steep $f$ dependence drives all systems rapidly though high frequencies. BHs therefore produce stronger strains at low frequencies, allowing them to dominate, while lighter binaries dominate higher frequencies by number.}

Whereas the signal is deterministic, the instrumental noise is stochastic, in that the strain time series of the noise will be a randomised realisation of the underlying process defined by the sensitivity curve $S_\mathrm{noise}(f)$. Thus, to simulate the PSD that LISA will actually detect, we simulate a random realisation of the instrumental noise. To do this, for each Fourier frequency $f_j$, we select random variables for the real and imaginary parts of $N(f_j)$ ( which is the discrete Fourier transform of $n(t_k)$) from a Gaussian distribution with a mean of zero and a variance of $S_\mathrm{noise}(f_j)/2$ \citep{OnGeneratingPowerLawNoise,2013MNRAS.434.1476I}. This procedure selects random phases for each Fourier frequency whilst ensuring that the noise PSD is equal to $S_\mathrm{noise}(f_j)$.

\begin{figure}
    \centering
    \includegraphics[width=\columnwidth]{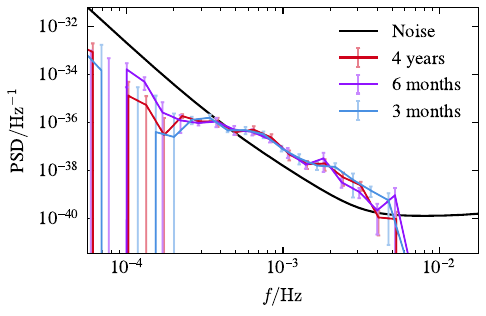}
    \vspace{-1.5\baselineskip}
    \caption{The LISA-detected binary signal, obtained by summing simulated instrumental noise and the total binary signal, then subtracting the theoretical noise, for different mission observation times. Even after just 3 months, the signal is already strong\rev{, and remains so because individually detectable binaries are not subtracted here.}}
    \label{fig:Signal_T_obs}
\end{figure}

We define the LISA-detected binary signal as the PSD obtained by summing the simulated instrumental noise with the total compact binary signal and subtracting the theoretical noise curve (the signal and noise can be summed because they are uncorrelated with each other; \citealt{1989ASIC..262...27V}). \hyperfig{fig:Signal_T_obs} shows the resulting PSD for $T_{\rm obs}=$ 3 months, 6 months, and 4 years. The apparent gaps in the PSD near $10^{-4}~\mathrm{Hz}$ arise from noise realisations that fall below the sensitivity curve, which become negative after subtraction when the signal is also weak. The error bars have increased due to the inclusion of instrumental noise realisations prior to geometric re-binning, with extreme values coming from the negative PSD values. Even after just 3 months, the LISA-detected binary signal is clearly present and detectable. \rev{This PSD does not decrease with observation time because here we retain all individually detectable sources; these grow more resolvable with time, and their iterative removal is carried out later in \hypersec{sec:resolvable_binaries}.} \edtai{We note, however, that we have assumed we can average over a full year of LISA's sky position and binary inclination (the factor of $\sqrt{4/5}$ in equation \ref{eq:amplitude_circular}), which is of course not possible in practice for less than a year of observing time. Whether the true signal is greater than or less than our estimate depends on the details of LISA's orbit in the early stages of the mission, which are not yet known.}



\subsection{Chirping binary systems}

An additional approximation we can refine is the assumption of orbital stability. So far, we have assumed binaries to be dynamically stable, with fixed orbital separations and periods. However, binaries lose orbital energy through emitting GWs, reducing their separation and increasing their orbital frequency until they merge. This is known as \textit{chirping}. Binaries with \rev{rapid orbital frequency evolution through the LISA band merge quickly} and therefore contribute little to the steady Galactic foreground.

\rev{We estimate the orbital frequency change over the observation time by evaluating the GW-driven evolution at the system's current parameters:
\begin{equation}
    \Delta f_{\rm orb} \approx \frac{df_{\rm orb}}{dt} ~ T_\mathrm{obs} = \frac{3}{2} \frac{f_{\rm orb}}{a} \left|\frac{da}{dt} \right| ~ T_\mathrm{obs}, 
\end{equation}
where $a$ is the semi-major axis, and $da/dt$ is the GW-driven orbital decay \citep[equation 5.6]{1964PhRv..136.1224P}. For almost all systems within our Compact binary population, $\Delta f_\mathrm{orb}$ is small enough that the corresponding GW frequencies remain within the same Fourier bins (of width $1/T_\mathrm{obs}$) over 4 years; only $<0.003\%$ of binaries near bin edges shift to a neighbouring bin. Thus, orbital frequencies evolve sufficiently slowly to be treated as stable over LISA's lifetime, and as such we neglect chirping in subsequent analysis. Furthermore, merger time estimates \citep[equation 5]{Mandel_2021} indicate that terminal GW frequencies remain well outside the LIGO band ($10$-$10^3~\mathrm{Hz}$) for the next $10^5$ years, consistent with the lack of observed Galactic compact binary mergers.}



\section{\edtai{Results}}
\label{sec:analyse_signal}

\subsection{The total PSD}

\begin{figure}
    \centering
    \includegraphics[width=\columnwidth]{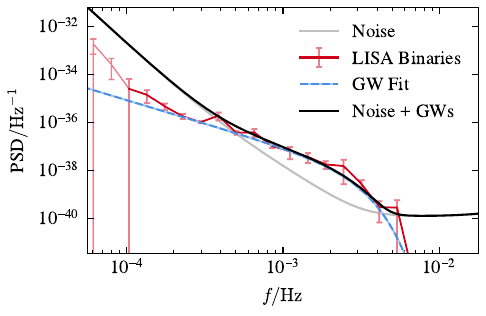}
    \vspace{-1.5\baselineskip}
    \caption{The LISA-detected binary signal, fit with an exponentially cut-off power law model (\hypereqalt{eq:power_law_exp_cutoff}). The lighter solid line shows the full LISA-detected binary signal, while the darker solid line highlights the frequency range used for the fit. The best fitting model is shown as a dashed line.}
    \label{fig:signal_fit}
\end{figure}

We first focus on the LISA-detected binary signal, which is the total noise-subtracted PSD including all of the Galactic binaries. Studying this signal provides a fast way to obtain meaningful early science results from the LISA mission\edita{, and we have made a simple \textsc{python} code to calculate such a PSD from any user-defined synthetic binary population publicly available (see the Data Availability section).} \hyperfig{fig:signal_fit} shows the LISA-detected PSD calculated for the default \textsc{cosmic} parameters and assuming 4 years of observation. To quantify the signal, we fit this PSD with an exponentially cut off power law function
\begin{equation}
    S = \mathcal{A}_\mathrm{c} ~ f^{-\gamma} ~ \exp\left[ - \left( f/f_\mathrm{c}\right)^2 \right].
    \label{eq:power_law_exp_cutoff}
\end{equation}
We fit this model across the frequency band $10^{-4}-10^{-2}~\mathrm{Hz}$. The best fit is shown as the dashed line in \hyperfig{fig:signal_fit}, and the best fitting parameters are $\mathcal{A}_\mathrm{c} =  1.4 \times 10^{-43} ~ \mathrm{Hz}^{-1}$, $\gamma = 2.0$, $f_\mathrm{c} =  2.3 \times 10^{-3}~\mathrm{Hz}$.

\begin{figure*}
    \centering
    \includegraphics[width=\textwidth]{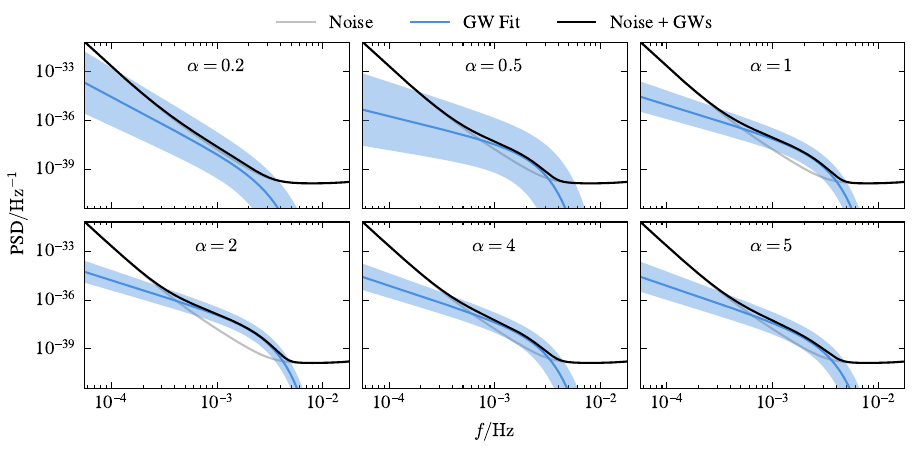}
    \vspace{-1.5\baselineskip}
    \caption{Fitted LISA-detected signal from Galactic binaries across different CE efficiency parameters ($\alpha$). Each panel shows the mean signal averaged over 50 realisations, with shaded regions representing $\pm1\sigma$ confidence intervals. As CE efficiency increases (left to right, top to bottom), the signal evolves significantly.}\label{fig:fit_panels}
\end{figure*}

\edtai{We now vary the CE efficiency parameter $\alpha$, and fit the above function to the LISA-detected PSD calculated for each $\alpha$ value. \hyperfig{fig:fit_panels} shows the best fitting model for each value of $\alpha$ trialled. Here, we have selected $50$ different realisations of Galactic positions for each $\alpha$ value (as described in Section \ref{sec:kicks}), and fit the exponential function to each PSD realisation. We then averaged the resulting fit parameters across realisations: the amplitude was geometrically averaged due to its spread over several orders of magnitude, while the cut-off frequency and power law index were arithmetically averaged. We take uncertainties (shaded regions represent $68\%$ confidence intervals) from the spread across different realisations, because these exceed individual fit errors.}

\edtai{We see that the shape and amplitude of the predicted PSD depends on the assumed $\alpha$ value.} At very low efficiency ($\alpha \lesssim 0.2$), the signal is faint and barely detectable above the instrumental noise (top left). As $\alpha$ increases to $ \sim 0.5$, a distinct bump emerges, though with significant spread between realisations (top middle). Between $0.5 \lesssim \alpha \lesssim 2$, the bump strengthens and its spread decreases, reaching maximum visibility near $\alpha \sim 2$ (top right and bottom left). At higher $\alpha$, the bump weakens again and the spread increases (bottom middle), and stabilises at very large $\alpha$ values (bottom right). \rev{The physical explanation for this $\alpha$ dependence is discussed in \hypersec{sec:discussion} and illustrated in \hyperfig{fig:CE_phase}.}

\begin{figure}
    \centering
    \includegraphics[width=\columnwidth]{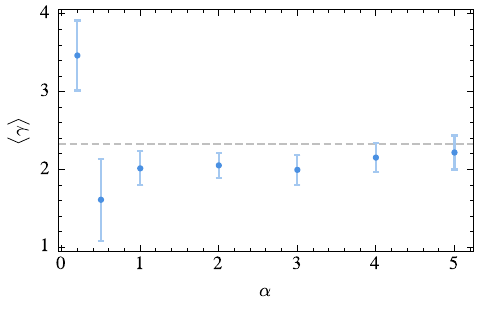}
    \vspace{-1.5\baselineskip}
    \caption{Mean power law index $\langle \gamma \rangle$ of the fitted LISA-detected PSD plotted against the CE efficiency parameter, $\alpha$. A sudden decrease is observed followed by a smooth increase towards $7/3$ (dashed), with the transition centred near $\alpha = 0.5$. Error bars indicate $\pm 1\sigma$ deviation from different realisations.}
    \label{fig:alpha_gamma}
\end{figure}

\hyperfig{fig:alpha_gamma} shows how the best fitting mean power law index, $\langle \gamma \rangle$, evolves with $\alpha$. We see that this parameter undergoes a sharp transition near $\alpha = 0.5$ \edita{(albeit with large uncertainties)}, dropping from above $3$ to below $2$, then rising smoothly towards a power law index of $7/3$. This final value corresponds to the theoretical slope expected for the GW PSD of a population of quasi-circular binaries \citep[equation 3]{2017JPhCS.840a2024C}. \edtai{Our results thus imply that it may be possible to distinguish between very low- and high-$\alpha$ regimes by studying the LISA PSD. We note that, although the results we present assume 4 years of observation, the analysis can be done much earlier in the LISA mission (since the measurement errors are smaller than the uncertainties introduced by the spread of different distance realisations).} The other two parameters are roughly constant with $\alpha$, with the amplitude $\langle  \mathcal{A_\mathrm{c}} \rangle \sim 10^{-44}~\mathrm{Hz}^{-1}$ and cut-off frequency $\langle f_\mathrm{c} \rangle \sim (1-3)~\mathrm{mHz}$.

\subsection{Individually detectable binaries}
\label{sec:resolvable_binaries}

We now consider \editj{individually detectable} binaries. We identify \editj{detectable} binaries from the SNR,
\begin{equation}
    \rho = \sqrt{\sum_{\eta = 1}^\infty \mathcal{A}_\eta^2 \frac{T_\mathrm{obs}}{S_\mathrm{eff}(f_\eta)}}, \label{eq:SNR_iterative}
\end{equation}
where $S_\mathrm{eff}$ is the sum of the simulated instrumental noise and the binary GW signal.
We iteratively identify \editj{detectable} systems by (1) calculating the effective signal approximated through geometric re-binning, (2) identifying and removing binaries with $\rho > 7$, and (3) repeating until no further binaries are \editj{individually detectable}. This results in a list of \editj{individually detectable} binaries and a residual PSD of unresolvable sources that contribute confusion noise.

\begin{figure}
    \centering
    \includegraphics[width=\columnwidth]{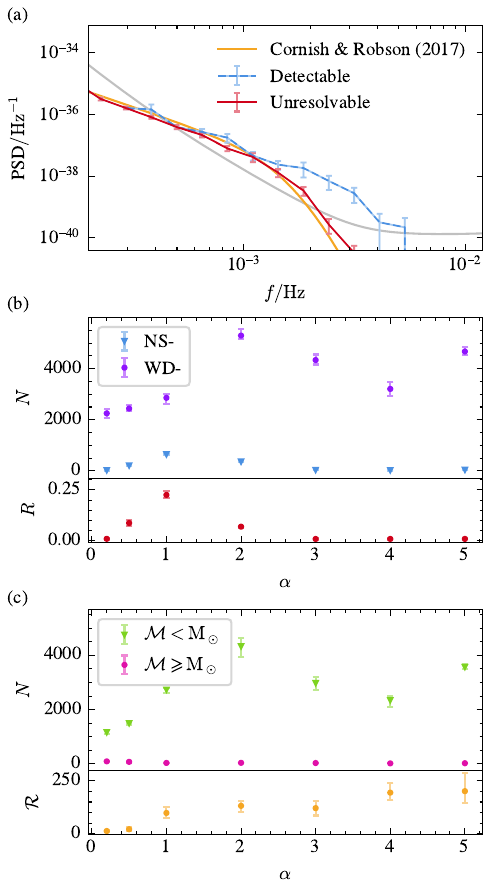}
    \vspace{-1.5\baselineskip}
    \caption{(a) Estimated confusion noise (solid line with error bars) from the unresolved Galactic binary population, obtained by filtering out \editj{individually detectable} sources from the total detectable signal (dashed line with error bars), \editj{both with noise subtracted. Importantly, the confusion noise agrees well with \citet{2017JPhCS.840a2024C}.} (b) Number of \editj{individually detectable} binaries as a function of the CE efficiency parameter, $\alpha$, separated into systems with NS and WD primaries. The lower panel shows the ratio $R$ of \editj{individually detectable} NS- to WD- binaries (triangular to circular markers), showing a clear trend with $\alpha$. (c) Number of \editj{individually detectable} binaries split by measured chirp mass, $\mathcal{M}$, above and below $1~\mathrm{M}_\odot$. The lower panel shows the ratio $\mathcal{R}$ of binaries with $\mathcal{M}<\mathrm{M}_\odot$ to those with $\mathcal{M}\geqslant \mathrm{M}_\odot$ (triangular to circular markers), which increases steadily with $\alpha$.}
    \label{fig:resolvable}
\end{figure}

\hypersubfig{fig:resolvable}{a} shows the \editj{noise-subtracted} unresolvable signal (solid line with error bars) calculated for the default \textsc{cosmic} parameters and assuming 4 years of observing time. We see that this lies below the total detectable signal (dashed line with error bars), and agrees well with the analytic confusion noise estimate of \citet{2017JPhCS.840a2024C}. \edtai{We find that the predicted confusion noise is not strongly sensitive to the assumed CE efficiency.} \hypertab{tab:resolvable} summarises the numbers of \editj{individually detectable} binaries that we predict for different values of $\alpha$. \edtai{Here, we have increased the numbers by the factor of $10$ described in \hypersec{sec:init_stellar_pop} to scale our simulated population up to the size of the entire Milky Way, and we have averaged our results across ten realisations of the distance model. \editj{Errors are given as asymmetric $1\sigma$ intervals corresponding to the 16th-84th percentile from the spread across these ten realisations. Where no uncertainty is shown, all realisations produced identical detections.}}

\hypersubfig{fig:resolvable}{b} shows how the number of \editj{individually detectable} binaries varies with $\alpha$, split into WD- and NS-primary binaries. While the absolute counts are uncertain, the ratio $R$ of \editj{individually detectable} NS- to WD- primary binaries is more robust and shows a clear trend: a symmetric peak near $\alpha = 1$ with $R \approx 0.25$, and lower values either side. \editj{The asymmetric uncertainties on $R$ are obtained by propagating the mean-count errors to first order; similar results are found when computing $R$ for each realisation and averaging, supporting its stability as a diagnostic.} This ratio therefore acts as an alternative diagnostic for $\alpha$, indicating whether the CE efficiency is near $1$ or in a low/high regime. By combining this with the power law slope $\langle \gamma \rangle$ from \hyperfig{fig:alpha_gamma}, we can in principle pinpoint the regime. If $\gamma > 7/3$, the value lies on the low-$\alpha$ side of the $R$ peak, whereas if $\gamma < 7/3$, the value lies on the high-$\alpha$ side.

In practice, identifying the source class of each binary may not always be observationally feasible. To address this, we also consider the measured chirp mass, $\mathcal{M}$, which provides a directly observable quantity to some uncertainty (e.g. see \citealt{2021MNRAS.502.5576K}). To simulate measuring $\mathcal{M}$, we draw a value for each binary from a Gaussian distribution centred on the true chirp mass, with a standard deviation estimated from \citet[equation 3]{2021MNRAS.502.5576K}. Here, rather than scaling the realisation numbers by the factor of 10 described in \hypersec{sec:init_stellar_pop}, we instead generate ten independent measurements per \editj{individually detectable} binary to represent observational sampling. However, only binaries with a fractional chirp mass uncertainty $\sigma_\mathcal{M}/\mathcal{M} < 1$ are retained. \hypersubfig{fig:resolvable}{c} shows the resulting number of \editj{individually detectable} binaries split by measured chirp mass above and below $\mathcal{M}=1~\mathrm{M}_\odot$. The ratio $\mathcal{R}$ of \editj{individually detectable} binaries with $\mathcal{M}<\mathrm{M}_\odot$ to those with $\mathcal{M}\geqslant \mathrm{M}_\odot$ increases monotonically with $\alpha$, providing a direct and practical observational probe of the CE efficiency. \editj{Similarly to $R$, the asymmetric uncertainties on $\mathcal{R}$ are obtained by propagating the mean-count errors to first order; comparable results are found when computing $\mathcal{R}$ for each realisation and averaging, supporting its stability as a diagnostic.} Together, these three statistical properties \rev{may} provide a powerful diagnostic to infer the underlying CE physics from LISA observations.  

\begin{table*}\renewcommand{\arraystretch}{1.4}
    \centering
    \caption{Summary of the estimated average number of \editj{individually detectable} binaries for each CE efficiency parameter, $\alpha$. Errors are given as asymmetric $1\sigma$ intervals \editj{corresponding to the 16th-84th percentile spread across ten realisations per each $\alpha$ value. Where no uncertainty is shown, all realisations produced identical detections.} The binaries are classified by their primary star: white dwarf (WD), neutron star (NS), or black hole (BH), with the secondary specified where relevant as main sequence (MS), stripped helium star main sequence (sMS), or another compact object.}\label{tab:resolvable}
    \begin{tabular}{lcccccccccccc} 
    \hline
    $\alpha$ & \textbf{WD-} & WD-MS & WD-sMS & WD-WD & \textbf{NS-} & NS-MS & NS-sMS & NS-WD & \textbf{BH-} & BH-MS & BH-NS & BH-BH\\
    \hline
    $0.2$ & $2250\substack{+165 \\ -169}$ & $1041\substack{+42 \\ -47}$ & -- & $1193\substack{+127 \\ -131}$ & $19\substack{+7 \\ -9}$ & $6\substack{+4 \\ -6}$ & $4\substack{+6 \\ -4}$ & $26\substack{+4 \\ -6}$ & $2\substack{+4 \\ -2}$ & -- & -- & $2\substack{+4 \\ -2}$ \\
    $0.5$ & $2442\substack{+130 \\ -101}$ & $156\substack{+21 \\ -16}$ & $700\substack{+31 \\ -12}$ & $1585\substack{+91 \\ -104}$ & $211\substack{+31 \\ -36}$ & $5\substack{+5 \\ -5}$ & $6\substack{+4 \\ -6}$ & $200\substack{+31 \\ -39}$ & -- & -- & -- & -- \\
    $1.0$ & $2860\substack{+161 \\ -246}$ & $35\substack{+36 \\ -35}$ & $502\substack{+59 \\ -65}$ & $2323\substack{+93 \\ -175}$ & $645\substack{+19 \\ -21}$ & -- & -- & $645\substack{+19 \\ -21}$ & $24\substack{+6 \\ -4}$ & -- & -- & $24\substack{+6 \\ -4}$ \\
    $2.0$ & $5298\substack{+273 \\ -122}$ & -- & $6\substack{+4 \\ -6}$ & $5292\substack{+269 \\ -120}$ & $362\substack{+16 \\ -20}$ & -- & -- & $362\substack{+16 \\ -20}$ & $2\substack{+4 \\ -2}$ & -- & -- & $2\substack{+4 \\ -2}$ \\
    $3.0$ & $4338\substack{+219 \\ -175}$ & $13\substack{+13 \\ -13}$ & $37\substack{+13 \\ -17}$ & $4287\substack{+203 \\ -168}$ & $33\substack{+17 \\ -13}$ & $16\substack{+10 \\ -6}$ & -- & $17\substack{+3 \\ -7}$ & $10$ & -- & $10$ & -- \\
    $4.0$ & $3207\substack{+269 \\ -271}$ & -- & $21\substack{+15 \\ -11}$ & $3186\substack{+253 \\ -263}$ & $25\substack{+15 \\ -15}$ & $10$ & -- & $15\substack{+15 \\ -15}$ & $20$ & $20$ & -- & -- \\
    $5.0$ & $4674\substack{+161 \\ -140}$ & -- & -- & $4674\substack{+161 \\ -140}$ & $37\substack{+9 \\ -7}$ & $14\substack{+6 \\ -4}$ & -- & $23\substack{+7 \\ -3}$ & $10$ & -- & -- & $10$ \\
    \hline
    \end{tabular}
\end{table*}

\section{Discussion} \label{sec:discussion}

\edtai{We have used binary population synthesis modelling to predict that it will be possible to detect the Galactic binary GW foreground early in the LISA mission by considering the total strain PSD. Early scientific inferences can already be made at that stage by fitting the PSD with an analytic function and comparing with the results of binary population models. Detection of individual sources will take longer, but will offer qualitatively new information. In particular, we find that both the total PSD and the relative numbers of \editj{individually detectable} binaries of different classes depend on the CE efficiency $\alpha$. We find that the power law index $\gamma$ characterising the PSD is steeper for the lowest values of $\alpha$, and that the ratio of NS primary systems to WD primary systems that are \editj{individually detectable} peaks at $\alpha\approx 1$ and drops off for higher and lower values. The latter is not a direct observable, since in real data we can not necessarily tell which source class each detection belongs to. We therefore also consider measured chirp mass, and find that the ratio of \editj{individually detectable} binaries detected with $\mathcal{M} < M_\odot$ to those detected with $\mathcal{M} \geqslant M_\odot$ increases with $\alpha$. By combining the statistical properties of the \editj{individually detectable} binaries and the results of fitting the total PSD, it should be possible to constrain the CE efficiency.} The CE phase is currently very poorly understood. Parameter estimates are as low as $\alpha \sim 0.2$ \citep{2010A&A...520A..86Z, 2022MNRAS.513.3587Z} to as high as $\alpha \sim 5$ \citep{2019ApJ...883L..45F}, and it has been shown that sources like X-ray binaries cannot be used to constrain CE models \citep{2014bsee.confE..37W}. Therefore this new diagnostic would be very valuable for our understanding of binary evolution.

\edtai{The physical reason for this $\alpha$ dependence is illustrated in \hyperfig{fig:CE_phase}}. For low $\alpha$, few binary systems survive the CE phase, resulting in fewer binaries in the present-day and therefore a weaker foreground GW signal. For very high $\alpha$, the CE phase is short lived as the envelope is ejected before the binary separation has reduced by much. Many binaries therefore survive the CE phase, but their separation is too large for them to be emitting GWs within the LISA bandpass in the present-day. At intermediate \rev{fiducial} values ($\alpha \sim 1$), there is a sweet spot where enough compact binaries survive with sufficiently tight orbits to produce a very prominent signal. This balance drives the evolution of the fitted power law index $\gamma$, which drops sharply at $\alpha = 0.5$ and then rises smoothly towards the theoretical value of $7/3$ for a population of circular binaries. The other two parameters of the PSD model, the cut-off frequency $f_\mathrm{c}$ and the amplitude $\mathcal{A}_c$, are consistent across all values of $\alpha$ that we trialled. If future observations yield a value of $\gamma$ or $f_\mathrm{c}$ significantly different from this prediction, it would indicate that our population synthesis assumptions must be revised.

\edtai{We use the \textsc{cosmic} population synthesis code to evolve a population of binaries accounting for a realistic Galactic SFH, and we predict the GW signal from each binary taking into account the effect of supernova kicks on the present-day position of binaries by evolving their post-kick positions in the Galactic potential (using the code \textsc{galpy}). \editj{The natal kick distributions employed here are the standard choice for population synthesis and related studies, however, the natal kick distribution of compact objects remains subject to much debate. There is growing evidence that BHs may receive high natal kicks, comparable to those applied to NSs (e.g. \citealt{2012MNRAS.425.2799R, 2024MNRAS.527L..82D}). One cannot easily quantify the impact that different kick magnitudes would have on the results; it may be that strong natal kicks result in fewer binaries hosting a BH, however disruption rates are not directly correlated to kick magnitudes (indeed, high natal kicks are thought to be favourable for double compact object mergers; \citealt{2016ApJ...819..108B}). \rev{Furthermore, \citet{2025ApJ...989L...8D} recently demonstrated that the widely used Hobbs Maxwellian distribution \citep{2005MNRAS.360..974H} may overestimate kick velocities when correctly fitted with an additional Jacobian, implying that our adopted kick prescriptions could be too extreme.}} 

\begin{figure}
    \centering
    \includegraphics[width=\columnwidth]{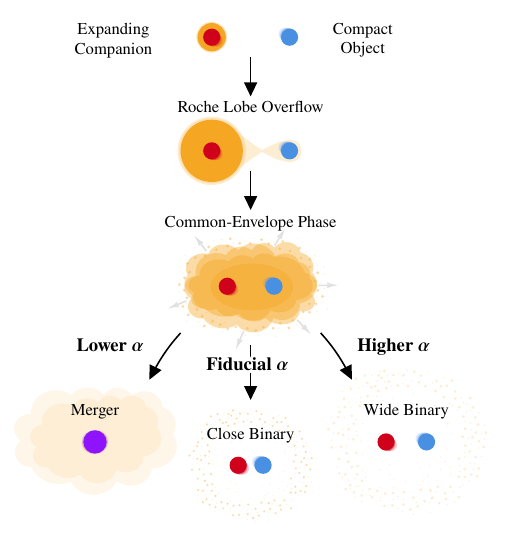}
    \vspace{-2\baselineskip}
    \caption{{Schematic of the common envelope (CE) phase in binary evolution. As the donor star evolves and its envelope expands, unstable mass transfer can cause the formation of a CE. This ultimately leads to mergers or a tighter binary pair.}}
    \label{fig:CE_phase}
\end{figure}

Considering all binaries with a BH, NS, or WD primary, we predict $\sim (2-5) \times 10^3$ binaries to be \editj{individually detectable} after a 4 year mission (see Table \ref{tab:resolvable}), with the number of WD primary systems roughly increasing with $\alpha$ and the number of NS primary systems peaking at $\alpha \sim 1$. Although our overall numbers are lower than some other studies \citep[e.g.][]{2017JPhCS.840a2024C}, the trends broadly agree with earlier work. This is unsurprising because the total normalisation is somewhat uncertain. Two assumptions contribute a simple scaling to our detection numbers. First, we normalise our population size assuming a constant binary fraction of $50\%$ (i.e. there are as many single stars as there are binary systems), but
\editj{in reality, the binary fraction of the Galaxy depends on stellar-mass and location \citep{2007ApJ...670..747K, 2024ApJ...964..164G}, increasing to $>80\%$ in the high mass regime.} Increasing the binary fraction to $100\%$ would increase our numbers by a factor of \editj{$\sim 1.64$ (i.e. the computed ratio of binaries at $100\%$ to $50\%$ for the same initial total stellar mass)}. We also scale our initial population to have a total stellar mass of $6.1 \times 10^{10}~M_\odot$, which is the total measured stellar mass in the Galaxy \textit{today} \citep{2015ApJ...806...96L}. \citet{Wagg+2022} estimate the initial stellar mass to be a factor $\sim 1.7$ larger than the present-day mass, since mass has been lost along the way e.g. to stellar winds. The numbers of predicted \editj{individually detectable} binaries we present here are therefore rather conservative, and could be potentially increased by a factor of roughly \editj{$1.64 \times 1.7 \approx 2.8$}.}

Our study considers a greater variety of binary classes than most contemporary studies, which have typically focused on a subset of classes for computational efficiency. It is therefore useful to compare our results on various binary subclasses to several other studies.
Our results on NS-WD systems are broadly consistent with \citet{2024MNRAS.530..844K}, with the caveat that direct comparison is made difficult by our use of a slightly different CE prescription \editj{(they assumed a constant $\lambda$ whereas we assume $\lambda$ to depend on evolutionary state and envelope mass following \citealt{2014A&A...563A..83C})}. For their Hobbs kick model (which aligns with our assumptions), they predict $\sim 100$ NS-WD binaries for $\alpha\lambda=0.25$ and $\alpha\lambda=2$. These two data points are roughly consistent with our results in \hypertab{tab:resolvable} if on average we have $\lambda \approx 0.7$ in our calculation. We predict the same rising trend in detected WD-WD systems with increasing $\alpha$ as reported by \citet{2024MNRAS.530..844K} and \citet{2025ApJ...981...66D}. Our absolute numbers are rather smaller, but within the normalisation uncertainty of $2.8$ discussed above. We predict that the number of \editj{individually detectable} WD-MS systems decreases with increasing $\alpha$ (see \hypertab{tab:resolvable}). \citet{2023MNRAS.525L..50S} predict that $\sim 100$ cataclysmic variable stars (CVs) will be detected, using a mock catalogue instead based on observed CVs \editj{with a space density broadly consistent with local estimates from our population ($\rho_0 \approx 5\times 10^{-6}~\mathrm{pc^{-3}}$ for $\alpha = 1$).} \editj{To match these existing observation-based predictions, we therefore should be predicting at least $100$ detected WD-MS systems. This is a conservative estimate, since not all systems will necessarily be accreting or as detectable as CVs, but it already favours $\alpha \lesssim 1$ based upon \hypertab{tab:resolvable}.}

\edtai{We predict only small numbers of \editj{individually detectable} BH-NS and BH-BH binaries, and no NS-NS systems (see \hypertab{tab:resolvable}). Our simulated population is too small for these numbers to be seen as reliable (after scaling by a factor of 10 to represent the true size of the Milky Way better, a prediction of 20 detections corresponds to just two detections from our sample, meaning that our population is too small to capture particularly rare eventualities), but based on our research we can state that they are small. \rev{These numbers are smaller than those predicted by \citet{Wagg+2022}, although consistent within the considerable uncertainties associated with predicting such rare objects. Part of this discrepancy reflects our conservative approach to normalising the population size: we adopt a $\sim 40\%$ lower total stellar mass than \citet{Wagg+2022} and a smaller binary fraction than the simulation datasets of \citet{2021MNRAS.508.5028B, 2022MNRAS.516.5737B}, which assume a binary fraction of $100\%$ and employ a different population synthesis code.} We do note, however, that \citet{Wagg+2022} do not account for the effect of supernova kicks on the present-day source position, meaning that they likely underestimate distances on average \editj{(as supernova kicks tend to redistribute systems away from their birth sites to further distances; see \hypersubfig{fig:Population_Distance}{a})} and thereby \rev{slightly} overestimate their SNR relative to the noise. We also predict small numbers \editj{($<20$; see \hypertab{tab:resolvable}) of individually detectable} NS-MS and BH-MS sources, which could potentially also be detected as X-ray binaries in future.}



\edtai{It is encouraging that the trends with $\alpha$ that we uncover here seem to be consistent with other studies, but the large variety in total numbers across studies highlights the dependence of predictions on a highly complex interplay of parameters and assumptions. \rev{Although our results show variations in $\alpha$ produce measurable changes in the LISA PSD, future larger parameter studies exploring a wider range of assumptions (e.g. mass transfer efficiency, which may produce similar effects) are needed before $\alpha$ can be uniquely constrained.} This will be highly computationally intensive, but may be enabled by training neural networks to act as fast drop-in replacements to complex population synthesis codes (e.g. such as the simulation based inference applied to isolated pulsars by \citealt{2024ApJ...968...16G}). There are also many assumptions that are simplifications of more complex physics. CE evolution is a good example, since we have simply parameterised our ignorance of a complex physical process with an efficiency parameter.
We also note that we have not considered triple systems, whereas most BH progenitors are thought to form in triple systems \citep{2012Sci...337..444S}, and thus triple interactions could provide a formation channel to creating ultra-compact GW sources involving BHs that our analysis cannot explore \citep{2025arXiv250813264X}. We have also ignored the effects of dynamical interactions\edita{, and we do not include globular clusters (in which dynamical interactions are important, \citealt{2017MNRAS.469.3088K}) in our population synthesis model}.

These caveats aside, LISA will provide novel constraints on the binary population that will be even more powerful when combined with other multi-messenger techniques. Whereas LISA is most sensitive to close binaries, direct astrometric observations of BH-MS binaries are most sensitive to wide binaries \citep{2018ApJ...861...21Y}. Searches for self gravitational lensing flares from $\sim$edge-on BH-MS and NS-MS binaries \citep[e.g.][]{2021MNRAS.507..374W} are instead biased to intermediate separations (wider binaries feature stronger gravitational lenses, but flares from them are seen for a narrower range of inclinations). Simultaneous modelling of all these populations, \rev{exploring a wide range range of simulation parameters}, in conjunction with X-ray binaries, CVs and so on, will provide the best constraints on our understanding of binary evolution.}

\section{Conclusions}
\label{sec:conclusion}

We have modelled the GW signal from a synthetic population of Galactic compact binaries and predicted how these signals will appear to LISA. \edtai{We show that the summed GW signal from the Galactic binary population will already be detectable 3 months into the LISA mission, via analysis of the PSD of the total GW strain. \edita{We provide a simple publicly available \textsc{python} code to calculate such a PSD (see the Data Availability section).} We find that the shape of the PSD depends on the CE efficiency parameter $\alpha$, such that very low values ($\alpha \lesssim 0.25$) can in principle be preferred or ruled out early in the mission. We find that the number of \editj{individually detectable} binaries over the full 4 year mission also depends on $\alpha$. In particular, we predict that the ratio of \editj{individually detectable} binaries with chirp mass $\mathcal{M} < M_\odot$ to those with $\mathcal{M} \geqslant M_\odot$ increases with $\alpha$. We therefore conclude that it will \rev{may} be possible to place novel constraints on the CE efficiency by combining inferences from the PSD shape and statistics of individually detectable binaries. We conservatively predict that, during its full 4-year lifetime LISA will individually detect thousands of binaries, $>2000$ with a WD primary, $\gtrsim 20$ with a NS primary, and $\sim (0-30)$ with a BH primary.}

\section*{Acknowledgements}

This work made use of the Rocket HPC facility at Newcastle University. J.M. acknowledges its service from 2017 to 2026, now retired. J.M. and A. Ingram acknowledge support from the Royal Society. A. Igoshev thanks the Royal Society for the University Research Fellowship URF/R1/241531. G.W. was supported by the Polish National Science Center (NCN) through the grants 2018/30/A/ST9/00050 and 2021/41/B/ST9/01191. M.M. acknowledges support via STFC small award ST/Y001699/1. S.S. acknowledges support by STFC grant ST/T000244/1 and ST/X001075/1. \rev{We thank the anonymous reviewer for their helpful comments, which improved this paper.}

\section*{Data Availability}

The simulated populations used in this work \edita{and a \textsc{python} code to generate a LISA noise-subtracted PSD from a given population} are publicly available at \url{https://github.com/JakMc/LISA_Binaries}.



\bibliographystyle{mnras}
\bibliography{ref} 



\appendix

\section{COSMIC Parameters}\label{appendix:cosmic_parameters}

The full list of parameters used for our \textsc{cosmic} runs is presented in Table \ref{tab:COSMIC_parameters}.

\begin{table*}
    \centering
    \caption{Summary of \textsc{cosmic} input parameters, with full descriptions available in the documentation of \citet{2025zndo..15164778C}.}
    \label{tab:COSMIC_parameters}
    \begin{tabular}{lclcc} 
    \hline
    Flag & Parameter & \qquad\qquad\quad Description & Type & Default \\
    \hline
    \textbf{Metallicity}	& \texttt{zsun} & Solar metallicity  & $Z_\odot$ & 0.014 \\

    \textbf{Wind} & \texttt{windflag}   & Stellar wind & \texttt{model} &\texttt{3} \\
    & \texttt{eddlimflag} & Metallicity dependence & \texttt{model} &\texttt{0}  \\
    & \texttt{neta} 	  & Reimers mass loss efficiency& $\eta $ & 0.5 \\
    & \texttt{bwind} 	  & Binary enhanced mass loss & $B_\mathrm{w}$ &0.0 \\
    & \texttt{hewind}     & Helium star mass loss factor & $(1-\mu)$ &0.5 \\
    & \texttt{beta}       &	Wind velocity & \texttt{model} &\texttt{-1} \\
    & \texttt{xi}         &	Wind accretion efficiency &	$\mu_\mathrm{w}$ &0.5 \\ 
    & \texttt{acc2}       & Bondi-Hoyle accretion factor &$\alpha_\mathrm{w}$ & 1.5 \\ 

    \textbf{Common Envelope} & \texttt{alpha1} & Common envelope efficiency & $\alpha$ & 1.0 \\
    & \texttt{lambdaf} & Envelope binding energy & \texttt{model} & \texttt{0} \\
    & \texttt{ceflag} &  Orbital energy & \texttt{model} & \texttt{1} \\
    & \texttt{cekickflag} & Supernova (SN) separation & \texttt{model} & \texttt{2} \\
    & \texttt{cemergeflag} & Forced merger & \texttt{model} & \texttt{1} \\
    & \texttt{cehestarflag} & Roche-lobe overflow (RLO)  & \texttt{model} & \texttt{0} \\
    & \texttt{qcflag} & Mass ratio stability & \texttt{model} & \texttt{1} \\
    & \texttt{qcrit\_array} & Critical mass ratios & $q_\mathrm{crit}$ & Critical mass ratios for each stellar type\\

    \textbf{Kicks} & \texttt{kickflag} & SN kick & \texttt{model}& \texttt{0} \\
    & \texttt{sigma} & Maxwellian velocity dispersion  & $\sigma/\mathrm{km\,s^{-1}}$ &  265.0 \\
    & \texttt{bhflag} & BH kick & \texttt{model}& \texttt{1} \\
    & \texttt{bhsigmafrac} & BH kick scaling & \texttt{model}& \texttt{1} \\
    & \texttt{sigmadiv} & Electron-capture SN (ECSN) kick & \texttt{model}  & \texttt{-20} \\
    & \texttt{ecsn} & Max He-star mass for ECSN & $M_\mathrm{ecs}^+/\mathrm{M_\odot}$ & $2.25$ \\
    & \texttt{ecsn\_mlow} & Min ECSN mass limit & $M_\mathrm{ecs}^-/\mathrm{M_\odot}$& $1.6$ \\
    & \texttt{aic} & Accretion induced SN kick & \texttt{model}& \texttt{1} \\
    & \texttt{ussn} & Ultra-stripped SN kick & \texttt{model}&  \texttt{1}\\
    & \texttt{pisn} & Pair-instability & \texttt{model}& \texttt{-2} \\
    & \texttt{polar\_kick\_angle} & SN kick polar opening angle & $\phi/^\circ$ & 90.0\\
    & \texttt{natal\_kick\_array} & Manual inputs for kicks & \texttt{model} & Custom SN kick parameters for both stars involved \\
    
     \textbf{Remnants} & \texttt{remnantflag} & NS/BH remnant & \texttt{model}& \texttt{4}\\
    & \texttt{mxns} & NS/BH mass boundary & $M_\mathrm{xNS}/\mathrm{M_\odot}$& 3.0\\
    & \texttt{rembar\_massloss} & Collapse mass loss to neutrinos & \texttt{model} & \texttt{1/2} \\
     & \texttt{wd\_mass\_lim} & Enforce Chandrasekhar limit & \texttt{model} & \texttt{1} \\
     & \texttt{bhspinflag} & BH spins & \texttt{model}& \texttt{0}\\
     & \texttt{bhspinmag} & BH spin magnitude & $a$ & 0.0 \\
    
    \textbf{GR Decay} & \texttt{grflag} & Gravitational wave orbital decay & \texttt{model} & \texttt{1} \\
     \textbf{Mass Transfer} & \texttt{eddfac} & Eddington limit & \texttt{model} & \texttt{1} \\
    & \texttt{gamma} & Angular momentum (RLO) & \texttt{model} & \texttt{-2} \\
     & \texttt{don\_lim} & Donor mass loss (RLO)  & \texttt{model} & \texttt{-1} \\
    & \texttt{acc\_lim} & Accretion rate limit (RLO) & \texttt{model} & \texttt{-1} \\
    
    \textbf{Tides} & \texttt{tflag} & Tidal circularisation &  \texttt{model} & \texttt{1}\\
    & \texttt{ST\_tide} & Tides  &  \texttt{model} & \texttt{1}\\
    & \texttt{fprimc\_array} & Convective tide efficiency & $f_\mathrm{conv}$ &Convective tide scaling factors for each stellar type\\
    
    \textbf{White Dwarfs} & \texttt{ifflag} & WD mass modification &  \texttt{model} & \texttt{0} \\
    & \texttt{wdflag} & WD cooling law &  \texttt{model} & \texttt{1} \\
    & \texttt{epsnov} & Nova mass capture fraction & $\varepsilon$ &0.001\\
    
    \textbf{Pulsar} & \texttt{bdecayfac} & Magnetic field decay &  \texttt{model} & \texttt{1} \\
    &  \texttt{bconst} & Field decay scaling & $k/\mathrm{Myr}$ &3000.0 \\
    &  \texttt{ck} & Field decay timescale & $\tau_\mathrm{B}/\mathrm{Myr}$ &1000.0 \\
    
     \textbf{Mixing} & \texttt{rejuv\_fac} & Main sequence (MS) merger mixing & $f_\mathrm{mix}$ &1.0 \\
     & \texttt{rejuvflag} & MS mixing & \texttt{model} & \texttt{0} \\
     & \texttt{bhms\_coll\_flag} & BH collisions & \texttt{model} & \texttt{0}\\
    
    \textbf{Breaking} & \texttt{htpmb} & Magnetic braking &\texttt{model} & \texttt{1} \\
    
    \textbf{Misc.} & \texttt{ST\_cr} & Convective/radiative boundary & \texttt{model} & \texttt{1} \\
    \hline
    \end{tabular}
\end{table*}

\section{Discrete Fourier Transforms}\label{appendix:DFT}

We use the following definition of the discrete Fourier transform 
\begin{align}
    H(f_j) &= 2 \sqrt{\frac{dt}{N}} \sum_{k=1}^N h(t_k) ~ \mathrm{e}^{2\pi i jk /N}, \\
    h(t_k) &= \frac{1}{2}\frac{1}{\sqrt{Ndt}} \sum_{j=-N/2+1}^{N/2} H(f_j) ~ \mathrm{e}^{-2\pi i jk /N}. \label{eqn:DFT_def_1}
\end{align}
Under this definition, Parseval's theorem is
\begin{equation}
    \sum_{j=1}^{N/2} |H(f_j)|^2~df = \frac{2}{N} \sum_{k=1}^N h^2(t_k),
    \label{eqn:parsevals}
\end{equation}
which states that the power integrated over all positive frequencies is equal to twice the variance of the corresponding time series (which is always real for our application). The variance of a cosine of amplitude $\mathcal{A}$ is $\mathcal{A}^2/2$. Therefore the PSD of a cosine time series integrated over all frequencies is equal to $\mathcal{A}^2$.


\bsp	
\label{lastpage}
\end{document}